\documentclass[a4paper]{jpconf}
\usepackage{graphicx}
\usepackage{bm}

\newcommand{\smGT}{{\scriptscriptstyle >}}
\newcommand{\smLT}{{\scriptscriptstyle <}}
\newcommand{\smD}{{\rm\scriptscriptstyle D}}
\newcommand{\smCoul}{{\rm\scriptscriptstyle coul}}

\newcommand{\footnoteskip}{\baselineskip 12pt plus 1pt minus 1pt}

\newcommand{\bibliographyskip}{\baselineskip16pt plus 1pt minus 1pt}

\begin{document}

\hfill LA-UR-07-5874

\title{Calculating the charged particle stopping power\\[-5pt]
exactly to leading and next-to-leading order}

\author{Robert~L Singleton Jr}
\address{Los Alamos National Laboratory, 
     Los Alamos, New Mexico 87545 USA}
\ead{bobs1@lanl.gov}

\begin{abstract}
  I will discuss a new method for calculating transport quantities,
  such as the charged particle stopping power, in a weakly to
  moderately coupled plasma. This method, called dimensional
  continuation, lies within the framework of convergent kinetic
  equations, and it is powerful enough to allow for systematic
  perturbative expansions in the plasma coupling constant. In
  particular, it provides an {\em exact} evaluation of the stopping
  power to leading and next-to-leading order in the plasma coupling,
  with the systematic error being of cubic order. Consequently, the
  calculation is near-exact for a weakly coupled plasma, and quite
  accurate for a moderately coupled plasma. The leading order term in
  this expansion has been known since the classic work of Spitzer. In
  contrast, the next-to-leading order term has been calculated only
  recently by Brown, Preston, and Singleton (BPS), using the
  aforementioned method, to account for {\em all} short- and
  long-distance physics accurate to second order in the plasma
  coupling, including an exact treatment of the quantum-to-classical
  scattering transition.  Under conditions relevant for inertial
  confinement fusion, BPS find the alpha particle range in the DT
  plasma to be about 30\% longer than typical model predictions in the
  literature.  Preliminary numerical studies suggest that this renders
  the ignition threshold proportionally higher, thereby having
  potential adverse implications for upcoming high energy density
  facilities. Since the key ideas behind the BPS calculation are
  possibly unfamiliar to plasma physicists, and the implications might
  be important, I will use this opportunity to explain the method in a
  pedagogical fashion.
\end{abstract}

\vskip-0.5cm 

For a weakly to moderately coupled plasma, the charged particle
stopping power $dE/dx$ was recently calculated from first principles
in Ref.~\cite{bps} using the method of dimensional
continuation~\cite{lfirst}.  While the calculational techniques were
imported from quantum field theory, the calculation itself lies
squarely within the standard framework of convergent kinetic
equations.  I will assume some familiarity with Ref.~\cite{bps},
although a much shorter and self-contained exposition can be consulted
in Ref.~\cite{bps1}.  For ease of presentation, I will work here with
a one component plasma of charge $e$, temperature $T$, and number
density $n$, although the work described above is in the context of a
general multi-component plasma.  Our starting point will be the fact
that any plasma quantity can be written as a series expansion in {\em
integer} powers of a dimensionless plasma parameter~\cite{by}.  For
example, the stopping power takes the form
\begin{eqnarray}
  \frac{dE}{dx} 
  = 
  -\underbrace{A\, g^2\ln g}_{\rm LO}
  \,+\, 
  \underbrace{~B\, g^2~}_{\rm NLO} 
  \,+\,  
  \underbrace{~{\cal O}(g^3)~}_{\rm error} ~ ,
  ~~
  {\rm with~coupling~parameter}
  ~~~
  g 
  &=&
  \frac{e^2\kappa_\smD}{T} \ ,
\label{dedtNLO}
\end{eqnarray}
where $\kappa_\smD^2=4\pi\,e^2\,n/T$.  I have indicated the leading
order in $g$ term (LO) and the next-to-leading order term (NLO) in
Eq.~(\ref{dedtNLO}), while the minus sign in the LO term is a
convention that renders $A$ positive when energy is transferred to the
plasma.  Note that nonanalytic terms such as $\ln g$ can also appear
in the $g$-expansion. For the process of energy exchange via Coulomb
interactions, this non-analyticity arises from the competition between
disparate physical length scales.  The coupling $g$ is just the ratio
of the Coulomb energy, for two particles a Debye length $\lambda_\smD=
\kappa_\smD^{-1}$ apart, to the thermal kinetic energy as measured by
$T$. To get a feel for the numbers, one finds $g = 0.04$ at the center
of the sun, while in the ICF ignition regime $g$ can be smaller. In
such cases, {\em provided} we can calculate $A$ and $B$,
expression~(\ref{dedtNLO}) therefore gives an accurate evaluation of
$dE/dx$.

The coefficient $A$ was first calculated by Spitzer some time ago,
while $B$ was recently calculated in Ref.~\cite{bps} using a
regularization technique from quantum field theory called dimensional
continuation.  It is convenient to define the dimensionless
coefficient $C$ by \hbox{$B=-A\, \ln C$}, along with $K = A\, g^2$,
and to then express the stopping power as
\begin{eqnarray}
  \frac{dE}{dx} 
  &=& 
  K\, \ln\Lambda_\smCoul \,+\, {\cal O}(g^3) \ ,
  ~\rm{with}~~
  \ln\Lambda_\smCoul = -\ln\!\left\{ C\, g\right\} \ .
\label{lngsqu}
\end{eqnarray}
We see, then, that knowing the next-to-leading order term is
equivalent to knowing the exact coefficient $C$ under the
logarithm. In the extreme classical and quantum limits, the
logarithm can be written as the ratio of two length scales
\begin{eqnarray}
  \ln\Lambda_\smCoul 
  = 
  \ln\!\left\{ \frac{b_{\rm max}}{b_{\rm min}}\right\} \ ,
\label{lnmodel}
\end{eqnarray}
where $b_{\rm max}$ is set by the Debye screening length
$\kappa_\smD^{-1}$, and the scale $b_{\rm min}$ is set either by the
distance of closest approach in the extreme classical limit, or the
thermal De Broglie length in the extreme quantum regime. In either
case, we find $b_{\rm min}/b_{\rm max} \propto g$, and we see that the
$g$-dependence in the Coulomb logarithm arises quite
naturally. Reference~\cite{bps} can therefore be thought of as a
calculation of the Coulomb logarithm, including the exact
interpolation between the extreme classical and quantum limits.

Let us now turn to convergent kinetic equations.  As suggested by
Ref.~\cite{hubbard}, one can view the Boltzmann and Leonard-Balescu
equations as providing complementary physics since they both succeed
and fail in complementary regimes. The Boltzmann equation (BE) gets
the short-distance physics correct, while the Leonard-Balescu equation
(LBE) captures the long-distance physics; conversely, the BE and the
LBE miss the long- and short-distance physics, respectively. This
complementarity motivates a class of kinetic equations of the
form~\cite{aono}
\begin{eqnarray}
  \frac{\partial f}{\partial t} + {\bf v} \! \cdot \!{\bm\nabla} f  
  = 
  B[f] + L[f] - R[f] \ ,
\label{convKE}
\end{eqnarray}
where $R[f]$ is a carefully chosen ``regulating kernel'' designed to
``subtract'' the long-distance divergence from the scattering kernel
$B[f]$ of the BE and the short-distance divergence from $L[f]$ of the
LBE. Furthermore, the kernel $R[f]$ must also preserve the correct
physics in the complementary regimes, namely, it must not damage the
correct short-distance physics of the BE and the correct long-distance
physics of the LBE. Reference~\cite{bps} can be viewed as a
systematic and rigorous implementation of this procedure, albeit in a
more abstract form, accurate to second order in $g$ (with more work,
one could systematically calculate to third and higher order in $g$).

We now examine how dimensional continuation regulates the kinetic
equations. For simplicity, we concentrate on the classical regime,
although to second order in $g$, quantum mechanics can be included by
using the quantum scattering amplitude in $B[f]$.  The classical BBGKY
hierarchy for the Coulomb potential is well defined and finite. We run
into divergences only when truncating the hierarchy to derive
lower-order kinetic equations, such as the Boltzmann and the
Lenard-Balescu equations. Interestingly, this truncation problem
occurs for the Coulomb potential, and only then in three spatial
dimensions $\nu=3$. Therefore, we can regulate the theory, rendering
it completely finite and well defined, by performing the integrals in
an arbitrary number of dimensions $\nu$. Upon using this procedure,
logarithmic divergences in three dimensions become finite simple poles
$1/(\nu-3)$ in arbitrary dimensions. One can then work entirely with
finite quantities. In the case of the stopping power, we find that the
long-distance pole from the BE exactly cancels the short-distance pole
from the LBE, and the result is therefore finite when we set $\nu=3$
at the end of the calculation.  This provides a finite and
well-defined result, obtained from a regularization prescription
constructed in a consistent fashion at {\em all length and energy
scales}.  This is a common and time honored regularization procedure
in quantum field theory, where it is called dimensional regularization.

I will now review some of the more salient features of the method. Let
${\bf x}_\nu$ and ${\bf v}_\nu$ denote the \hbox{$\nu$-dimensional}
position and velocity vectors of a particle.  The Coulomb potential
for two particles separated a distance $r=\vert {\bf x}_\nu -{\bf
x}_\nu^\prime\vert $ is $V_\nu(r)=C_\nu\, e^2/r^{\nu-2}$, where
$C_\nu=\Gamma(\nu/2-1)/ 4\pi^{\nu/2}$ is a spatially dependent
geometric factor.\footnote{
\footnoteskip
  See Ref.~\cite{bps1} for more details.
}
The distribution function $f_\nu$ will be defined so that $f_\nu({\bf
x}_\nu,{\bf v}_\nu,t)\, d^\nu x \,d^\nu v$ gives the number of
particles in a small hyper-volume $d^\nu x$ about ${\bf x}_\nu$ and
$d^\nu v$ about ${\bf v}_\nu$ at time $t$.  We can define
multi-point correlation functions in a similar manner, and in this way
we can construct the BBGKY hierarchy in an arbitrary number of
dimensions.  In dimensions $\nu>3$, the standard textbook derivation
of the BE goes through without an infrared divergent scattering kernel
$B_\nu[f]$. Furthermore, since the $\nu$-dimensional Coulomb potential
$V_\nu(r) \propto 1/r^{\nu-2}$ emphasizes short-distance over
long-distance physics when $\nu>3$, the BBGKY hierarchy reduces to the
Boltzmann equation to {\em leading} order in $g$ in these dimensions:
\begin{eqnarray}
  {\rm BBGKY} \Rightarrow
  \frac{\partial f_\nu}{\partial t} + 
  {\bf v}_\nu \! \cdot \!{\bm\nabla}_{\!x}\,  f_\nu 
  =
  B_\nu[f] 
  ~~~{\rm to~LO~in~} g {\rm ~for~}\nu > 3\ .
\label{BEsimpnu}
\end{eqnarray}
Here, the $\nu$-dimensional spatial gradient has been denoted by
${\bm\nabla}_{\!x}$.  Conversely, in dimensions $\nu<3$, the Coulomb
potential $V_\nu(r)$ emphasizes long-distance physics over
short-distance effects, and consequently, to leading order in $g$, the
BBGKY hierarchy reduces to the Lenard-Balescu equation in this spacial
regime:
\begin{eqnarray}
  {\rm BBGKY} \Rightarrow
  \frac{\partial f_\nu}{\partial t} + 
  {\bf v}_\nu \! \cdot \!{\bm\nabla}_{\!x}\, f_\nu 
  =
  L_\nu[f]
  ~~~{\rm to~LO~in~} g {\rm ~for~}\nu < 3 \ ,
\label{LBEsimpnu}
\end{eqnarray}
where the scattering kernel in the LBE is $L_\nu[f]$. Space does not
permit us to write down the exact forms of $B_\nu[f]$ and $L_\nu[f]$
here, but one may consult Ref.~\cite{bps} for the expressions. These
kinetic equations allow one to calculate the stopping power in $\nu>3$
and $\nu<3$, the results of which are presented in Sections~8 and 7 of
Ref.~\cite{bps}, respectively. The calculation involves performing a
series of momentum and wave number integrals in arbitrary dimensions
$\nu$, and reduces to the form
\begin{eqnarray}
  \frac{dE^\smGT}{dx}
  &=& 
  H(\nu)\,\frac{g^2}{\nu-3} 
  +
  {\cal O}(\nu-3) 
  \hskip0.66cm:~  {\rm LO~in}~g~{\rm when~}\nu > 3 \ ,
\label{dedtonecal}
\\[5pt]
  \frac{dE^\smLT}{dx}
  &=&
  G(\nu)\, \frac{g^{\nu-1}}{3-\nu} 
  + {\cal O}(3-\nu) 
  \hskip0.7cm :~ {\rm LO~in}~g~{\rm when~} \nu < 3 \ .
\label{dedttwocal}
\end{eqnarray}
The analytic expressions for $H(\nu)$ and $G(\nu)$ are rather
complicated,\footnote{
\footnoteskip
  For the related process of electron-ion temperature equilibration,
  in contrast, the expressions for $H(\nu)$ and $G(\nu)$ are quite
  simple. 
}
and space does not permit their reproduction here. In this paper, we
are only interested in their analytic properties as a function of
$\nu$. In particular, the coefficients $H(\nu)$ and $G(\nu)$ can be
expanded in powers of $\epsilon=\nu-3$, and we find
\begin{eqnarray}
  H(\nu) 
  =
  -A + \epsilon \,H_1 + {\cal O}(\epsilon^2)
  ~~~{\rm and}~~~
  G(\nu) 
  =
  -A + \epsilon \,G_1 + {\cal O}(\epsilon^2) \ .
\label{Gexp}
\end{eqnarray}
For our purposes, we do not require the exact forms of $H_1$, $G_1$,
nor that of the leading term~$A$. It is sufficient to note that the
leading terms in Eq.~(\ref{Gexp}) are equal, so that $H(\nu \equiv
3)=G(\nu \equiv 3)$.  This is a fact that arises from the calculation
itself, as it must, and it should be emphasized that this equality is
not arbitrarily imposed by hand.  It is a {\em crucial} point that the
leading terms are identical, as this will allow the short- and
long-distance poles to cancel, thereby giving a finite result.

Since the rates $dE^\smGT/dx$ of Eq.~(\ref{dedtonecal}) and
$dE^\smLT/dx$ of Eq.~(\ref{dedttwocal}) were calculated in mutually
exclusive dimensional regimes, one might think that they cannot be
compared. However, even though Eq.~(\ref{dedttwocal}) was originally
calculated in $\nu<3$ for integer values of $\nu$, we can analytically
continue\footnote{
\footnoteskip
  In the same way that the factorial function $n!$ on the positive 
  integers can be generalized to the Gamma function $\Gamma(z)$ over
  the complex plane, including both the positive and negative real axes. 
}  
the quantity $dE^\smLT/dx$ (viewed as a function of dimension $\nu$)
to real values of $\nu$ with $\nu>3$.  We can then directly compare
Eqs.~(\ref{dedtonecal}) and (\ref{dedttwocal}). Upon writing the
$g$-dependence of Eq.~(\ref{dedttwocal}) as $g^{2 + (\nu-3)}$, when
$\nu>3$ we see that Eq.~(\ref{dedttwocal}) is indeed higher order in
$g$ than Eq.~(\ref{dedtonecal}):
\begin{eqnarray}
  \frac{dE^\smLT}{dt}
  &=&
  -G(\nu)\, \frac{g^{2+(\nu-3)}}{\nu-3} 
  +
  {\cal O}(\nu-3) 
  ~~:~{\rm NLO~in}~g~{\rm when~} \nu > 3 \ .
\label{NLOgterm}
\end{eqnarray}
By power counting arguments, no powers of $g$ between $g^2$ and
$g^{\nu-1}$ can occur in Eq.~(\ref{dedtonecal}) for $\nu>3$, and
therefore Eq.~(\ref{dedttwocal}) indeed provides the correct
next-to-leading order term in $g$ when the dimension is analytically
continued to $\nu>3$, The individual pole-terms in
Eqs.~(\ref{dedtonecal}) and (\ref{NLOgterm}) will cancel giving a
finite result when the leading and next-to-leading order terms are
added. The resulting finite quantity will therefore be accurate to
leading and next-to-leading order in $g$ as the $\nu \to 3$ limit is
taken:
\begin{eqnarray}
  \frac{dE}{dx}
  =
  \lim_{\nu \to 3^+}
  \Bigg[
  \underbrace{~\frac{dE^\smGT}{dx}~}_{\rm LO}
  +
  \underbrace{~\frac{dE^\smLT}{dx}~}_{\rm NLO}
  \Bigg] + {\cal O}(g^3) \ .
\label{dedtorderg3}
\end{eqnarray}
Note that this does not lead to any form of ``double counting''
since we are merely adding the next-to-leading order term
(\ref{NLOgterm}) to the leading order term (\ref{dedtonecal}) at a
common value of~\hbox{$ \nu > 3$}.
We are now in a position to evaluate the limit in
Eq.~(\ref{dedtorderg3}). Defining $\epsilon=\nu-3$ as before,
note that $g^{\epsilon} = \exp\{\epsilon \ln g\} = 
1 + \epsilon\ln g + {\cal O}(\epsilon^2)$, 
which gives the relation
\begin{eqnarray}
  \frac{g^\epsilon}{\epsilon}
  = 
  \frac{1}{\epsilon} 
  +
  \ln g + {\cal O}(\epsilon) \ .
\label{gexp}
\end{eqnarray}
Substituting Eq.~(\ref{gexp}) into Eq.~(\ref{NLOgterm}),
adding this result to Eq.~(\ref{dedtonecal}), and then
taking the limit gives 
\begin{eqnarray}
  \frac{dE}{dx}
  =
  - A\, g^2 \ln g   +  B\, g^2  + {\cal O}(g^3) \ ,
\end{eqnarray}
with $B=H_1-G_1$, in agreement with Eq.~(\ref{dedtNLO}).  In this way,
BPS has calculated the charged particle stopping power accurate to
leading order and next-to-leading order in $g$.

\end{document}